\documentstyle[amssymb,twocolumn,aps,prd]{revtex}

\tightenlines

\begin{document}

\twocolumn[\hsize\textwidth\columnwidth\hsize\csname
@twocolumnfalse\endcsname
\title{Remarks on topological models and fractional statistics}
\author{\bf C. A. S. Almeida\thanks{E-mail address: carlos@fisica.ufc.br}}
\address{{\normalsize {\it Universidade Federal do
Cear\'{a} - Departamento de F\'{\i}sica }}\\ {\normalsize {\it
C.P. 6030, 60470-455 Fortaleza-Ce , Brazil }}}

\maketitle

\begin{abstract}

One of the most intriguing aspects of Chern-Simons-type topological models is the
fractional statistics of point particles which has been shown essential for our
understanding of the fractional quantum Hall effects. Furthermore these ideas are
applied to the study of high Tc superconductivity. We present here an recently
proposed model for fractional spin with the Pauli term.

On the other hand, in D=4 space-time, a Schwarz-type topological gauge theory with
antisymmetric tensor gauge field, namely $B\wedge F$ model, is reviewed.
Antisymmetric tensor fields are conjectured as mediator of string interaction.
A dimensional reduction of the previous model provides a (2+1) dimensional topological
theory, which involves a 2-form $B$ and a 0-form $\phi$. Some recent results
on this model are reported.

Recently, there have been thoughts of generalizing unusual statistics to extended
objects in others space-time dimensions, and in particular to the case of strings
in four dimensions. In this context, discussions about fractional spin and
antisymmetric tensor field are presented.
\end{abstract}

\vskip0.5cm
PACS\, 11.15.-q, 11.10.Kk,12.90.+b \vskip0.5cm

\vspace{3ex} ] 

\section{$B\wedge F$ Models.}

Schwarz-type theories are purely topological in the sense that their
partition functions are independent of the metric and that the only
observables in these theories are topological invariants of the underlying
spacetime manifold ${\cal M}$. Other observables describe linking and
intersection number of manifolds of any dimension.

Commonly called BF systems, they are characterized by a BRST-gauge fixed
quantum action which differ from the classical action only by a
BRST-commutator which contains the whole metric dependence of the quantum
action. On the other hand, since the vaccum expectation value of a
BRST-commutator vanishes, these field theories may be obtained from the
classical actions \cite{Birmingham:1991ty}. Furthemore, if we denote as $%
{\cal Q}$ the BRST-operator which is nilpotent, in these theories the
energy-momentum tensor is ${\cal Q} $ trivial, i.e.,
\begin{equation}
T_{\mu \nu }=\{{\cal Q} , \Phi _{\mu \nu }\}  \label{T}
\end{equation}
where $\Phi _{\mu \nu }$ represents fields and the metric.

Connected to BF systems, it is worth mentioning that antisymmetric tensor
fields theories have been studied during the past years. They play an
important role in the realization of the various strong-weak coupling
dualities among string theories. An antisymmetric tensor of rank $p-1$
couples naturally to an elementary extended object of dimension $p-2$,
namely a $(p-2)$ brane. As an example of an abelian BF system consider the
following metric independent action on an D-dimensional manifold ${\cal M}$.
\begin{equation}
S(D,p)=\int_{{\cal M}}B_p\wedge dA_{D-p-1},  \label{BFD}
\end{equation}
where $A$ and $B$ are forms , $p$ denoting their rank, $\wedge $ denoting
their wedge product and $d$ is the exterior derivative.

In particular the abelian $B\wedge F$ four-dimensional action is
\begin{equation}
S_{BF}=\int_{M_4}\left\{ B\wedge F\right\} .  \label{BF4}
\end{equation}

\begin{equation}
B=B_{\mu \nu }dx^\mu \wedge dx^\nu , F=dA, A=A_\mu dx^\mu .  \label{BA}
\end{equation}

This action is formulated in terms of the two-form potential $B$ while $F=dA$
is the field-strength of a one-form gauge potential $A$.

\vskip0.3cm

{\bf {Applications:}}

\vskip0.3cm

{\bf {$\bullet $}} Field theories describing the low-energy limit of
fundamental string theories typically contain higher-rank tensor fields.

{\bf {$\bullet $}} The topological contribution coming from $BF$ theories
appear even in those physical theories with non trivial physical Hamiltonian
where the $BF$ term appears as an interaction term.

{\bf {$\bullet $}} Color confinement models.

{\bf {$\bullet $}} Axionic cosmic strings.

{\bf {$\bullet $}} QCD strings.

{\bf {$\bullet $}} Topologically massive models.

\section{GAUGE INVARIANT MASSIVE $B\wedge F$ MODEL IN $D=4$.}

Our starting point is an abelian gauge theory which contains the vector
field $A$ and the antisymmetric field $B$, and incorporated the topological
term $B\wedge F$ in the four-dimensional action \cite{Allen:1991gb}
\begin{equation}
S_{BF}=\int_{M_4}\left\{ \frac 12H\wedge \,^{*}H-\frac 12F\wedge
\,^{*}F+kB\wedge F\right\} .  \label{bfm01}
\end{equation}

Here $H=dB$ is the field-strength of a two-form gauge potential $B$, $k$ is
a mass parameter, and $*$ is the Hodge star (duality) operator. The action
above is invariant under the following transformations:
\begin{equation}
\delta A=d\theta ,\delta B=d\Lambda ,  \label{2.4}
\end{equation}
where $\theta $ and $\Lambda $ are zero and one-form transformation
parameters respectively, and gives the equations of motion
\begin{equation}
d\,^{*}H=\kappa F  \label{2.7}
\end{equation}
and
\begin{equation}
d\,^{*}F=\kappa H.  \label{2.8}
\end{equation}

Applying $d\,^{*}$ on both sides of eq. (\ref{2.8}) and using the eq. (\ref
{2.7}), we get
\begin{equation}
(d^{*}d^{*}+\kappa ^2)F=0.  \label{gmt05}
\end{equation}

Repeating the procedure above in reverse order, we obtain the equation of
motion for $H$
\begin{equation}
(d^{*}d^{*}+\kappa ^2)H=0.  \label{gmt06}
\end{equation}

These equations can be rewritten as
\begin{equation}
(\square +\kappa ^2)F=0  \label{2.11}
\end{equation}
and
\begin{equation}
(\square +\kappa ^2)H=0.  \label{2.12}
\end{equation}

\vskip1.0cm


\section{ABELIAN GAUGE INVARIANT MASSIVE MODELS IN $D=3.$}


{\bf {$\bullet $ Dimensional reduction $\rightarrow $ $B\wedge \varphi $
models.}} \vskip0.3cm

Dimensional reduction is usually done by expanding the fields in normal
modes corresponding to the compactified extra dimensions, and integrating
out the extra dimensions. This approach is very useful in dual models and
superstrings. Here, however, we only consider the fields in higher
dimensions to be independent of the extra dimensions.

In this case, we assume that our fields are independent of the extra
coordinate $x_3.$ From (\ref{BF4}), on performing dimensional reduction as
described above, we get in three dimensions
\begin{equation}
S=\int_{M_3}\left\{ B\wedge d\phi +V\wedge F\right\} ,  \label{bfm03}
\end{equation}
where $V$ and $\phi $ are a 1-form and a 0-form fields respectively.

We recognize that $B\wedge d\phi $ is topological in the sense that there is
no explicit dependence on the space-time metric. One has to stress that this
term may not be confused with the two-dimensional version of the $B\wedge F$%
, which involves a scalar and a one-form fields. Moreover, a term that is
equivalent to the four-dimensional $B\wedge F$ term is present in action (%
\ref{bfm03}) (the so-called mixed Chern-Simons term, $V\wedge F$). \vskip%
0.3cm

{\bf {$\bullet $ Non-Chern-Simons gauge invariant massive models in $D=3.$}}

\vskip0.3cm

Now, in order to show the topological mass generation for the vector and
tensor fields, we consider the model with the topological term $B\wedge
d\phi $, and with propagation for the two-form gauge potential $B$ and for
the zero-form field, represented by the action
\begin{equation}
S=\int_{M_3}\left\{ \frac 12H\wedge \,^{*}H+\frac 12d\phi \wedge \,^{*}d\phi
+\kappa B\wedge d\phi \right\} ,  \label{gmt01}
\end{equation}
where the second term is a Klein-Gordon term, $\kappa $ is a mass parameter
and $H=dB$ is a three-form field-strength of $B$.

The action above is invariant under the following transformations:
\begin{equation}
\delta A=-d\theta ,\delta \varphi =\theta ,\delta B=d\Lambda ,
\end{equation}
where $\theta $ and $\Lambda $ are zero and one-form transformation
parameters respectively.

We follow here the same steps that has been used by Allen {\it et al.} \cite
{Allen:1991gb} in order to show the topological mass generation in the
context of $B\wedge F$ model. Thus, we find the equations of motion for
scalar and tensor fields, which are respectively $d\,^{*}H=\kappa d\phi $
and $d\,^{*}d\phi =-\kappa H. $ Consequently, we obtain the equations $%
(d^{*}d^{*}+\kappa ^2)d\phi =0$ and $(d^{*}d^{*}+\kappa ^2)H=0.$

These equations can be rewritten as
\begin{equation}
(\square +\kappa ^2)\partial _\mu \phi =0  \label{gmt061}
\end{equation}
and
\begin{equation}
(\square +\kappa ^2)H=0.  \label{gmt062}
\end{equation}

Therefore, the fluctuations of $\phi $ and $H$ are massive. Obviously, these
two possibilities can not occurs simultaneously. Indeed, in the most
interesting case, the degree of freedom of the massless $\phi $ field is
''eaten up'' by the gauge field $B$ to become massive and the $\phi $ field
completely decouples from the theory \cite{Medeiros:1999df}.

\vskip0.5cm

\section{$N=1 - D=4$ MASSIVE $B\wedge F$ $\rightarrow$ $N=2 - D=3$ MASSIVE $%
B\wedge \varphi$ MODELS.}


\vskip0.3cm

\begin{itemize}
\item  {\bf {$N=1-D=4$ massive $B\wedge F$ model.}} \vskip0.5cm
\end{itemize}

Let us begin by introducing the $N=1-D=4$ supersymmetric BF extended model.
For extended we mean that we include mass terms for the Kalb-Ramond field.
This mass term will be introduced here for later comparison to the
tridimensional case. Actually, this construction can be seen as a superspace
and abelian version of the so called BF-Yang-Mills models. These models are
described by the action
\begin{equation}
S_{BF-YM}=\int_{M_4}Tr\left\{ kB\wedge F+\frac{g^2}4B\wedge \,^{*}B\right\} .
\label{0.1}
\end{equation}

Note that, on-shell, (\ref{0.1}) is equivalent to the standard YM action.
This formalism was used by Fucito {\it et al.} \cite{Fucito:1997ax} in order
to study quark confinement.

As our basic superfield action we take \cite{Gomes:2001fr}
\begin{eqnarray}
S_{BF}^{SS} &=&\frac 18\int d^4x\{-i\kappa [\int d^2\theta B^\alpha
W_\alpha-\int d^2\overline{\theta }\overline{B}_{\dot{\alpha}}\overline{W}^{%
\dot{\alpha}}]  \nonumber \\
&&+\frac{g^2}2[\int d^2\theta B^\alpha B_\alpha +\int d^2\overline{\theta }%
\overline{B}_{\dot{\alpha}}\overline{B}^{\dot{\alpha}}]\}\mbox{ }.
\label{1.1}
\end{eqnarray}
where $W_\alpha $ is a spinor superfield-strength, $B_\alpha $ is a chiral
spinor superfield, $\bar{D}_{\dot{\alpha}}B_\alpha =0$, $\kappa $ and $g$
are massive parameters. Their corresponding $\theta $-expansions are:
\begin{eqnarray}
W_\alpha (x,\theta ,\bar{\theta}) &=&4i\lambda _\alpha (x)-[4\delta _\alpha
^\beta D(x)+2i(\sigma ^\mu \bar{\sigma}^\nu )_\alpha ^\beta F_{\mu \nu
}(x)]\theta _\beta  \nonumber \\
&&+4\theta ^2\sigma _{\alpha \dot{\alpha}}^\mu \partial _\mu \bar{\lambda}^{%
\dot{\alpha}}  \label{1.2a}
\end{eqnarray}
\begin{equation}
B_\alpha (x,\theta ,\overline{\theta })=e^{i\theta \sigma ^\mu \overline{%
\theta }\partial _\mu }[i\psi _\alpha (x)+\theta ^\beta T_{\alpha \beta
}(x)+\theta \theta \xi _\alpha (x)]\mbox{ ,}  \label{1.2b}
\end{equation}
where
\begin{equation}
T_{\alpha \beta }=T_{(\alpha \beta )}+T_{[\alpha \beta ]}=-4i(\sigma ^{\mu
\nu })_{\alpha \beta }B_{\mu \nu }+2\varepsilon _{\alpha \beta }(M+iN)\mbox{
}.  \label{1.2c}
\end{equation}

Our conventions for supersymmetric covariant derivatives are
\begin{eqnarray}
D_\alpha &\equiv &\frac \partial {\partial \theta ^\alpha }+i\sigma _{\alpha
\dot{\alpha}}^\mu \bar{\theta}^{\dot{\alpha}}\partial _\mu  \nonumber \\
\bar{D}_{\dot{\alpha}} &\equiv &-\frac \partial {\partial \bar{\theta}^{\dot{%
\alpha}}}-i\theta ^\alpha \sigma _{\alpha \dot{\alpha}}^\mu \partial _\mu %
\mbox{ .}  \label{1.2d}
\end{eqnarray}

We call attention for the electromagnetic field-strength and the
antisymmetric gauge field which are contained in $W_\alpha $ and $B_\alpha $%
, respectively. In terms of the components fields, the action (\ref{1.1})
can be read as
\begin{eqnarray}
S=\int d^4x\{[-\frac{i\kappa }2\left( \xi \lambda -\bar{\xi}\bar{\lambda}%
\right) +\frac \kappa 2B^{\mu \nu }\widetilde{F}_{\mu \nu }-\kappa DN]
\nonumber \\
+\frac \kappa 2\left( \psi ^\alpha \sigma _{\alpha \dot{\alpha}}^\mu
\partial _\mu \bar{\lambda}^{\dot{\alpha}}+\bar{\psi}_{\dot{\alpha}}\left(
\bar{\sigma}^\mu \right) ^{\dot{\alpha}\alpha }\partial _\mu \lambda _\alpha
\right)  \nonumber \\
+g^2[\frac 18\left( \psi \xi +\bar{\psi}\bar{\xi}\right) +\frac 12B^{\mu \nu
}B_{\mu \nu }-\frac 12\left( M^2+N^2\right) ]\}  \nonumber \\
=\int d^4x[(\frac{i\kappa }2\bar{\Xi}\gamma ^5\Lambda +\frac \kappa 2\bar{%
\Psi}\gamma ^\mu \partial _\mu \Lambda +\frac \kappa 2B^{\mu \nu }\widetilde{%
F}_{\mu \nu }-\kappa DN)  \nonumber \\
+g^2(\frac 18\bar{\Psi}\Xi +\frac 12B^{\mu \nu }B_{\mu \nu }-\frac 12\left(
M^2+N^2\right) )]\mbox{ }.  \label{comp}
\end{eqnarray}
In the last equality above, the fermionic fields have been organized as
four-component Majorana spinors as follows
\begin{equation}
\Xi =\left(
\begin{array}{c}
\xi _\alpha \\
\bar{\xi}^{\dot{\alpha}}
\end{array}
\right) \mbox{ };\mbox{ }\Lambda =\left(
\begin{array}{c}
\lambda _\alpha \\
\bar{\lambda}^{\dot{\alpha}}
\end{array}
\right) \mbox{ };\mbox{ }\Psi =\left(
\begin{array}{c}
\psi _\alpha \\
\bar{\psi}^{\dot{\alpha}}
\end{array}
\right) \mbox{ },  \label{1.4}
\end{equation}
and we denote the dual field-strength defining $\widetilde{F}_{\mu \nu
}\equiv \frac 12\varepsilon _{\mu \nu \alpha \beta }F^{\alpha \beta }$.

Furthermore, we use the following identities
\begin{eqnarray}
\bar{\Psi}\Lambda &=&\bar{\psi}\bar{\lambda}+\psi \lambda  \nonumber \\
\bar{\Psi}\gamma ^5\Lambda &=&\bar{\psi}\bar{\lambda}-\psi \lambda  \nonumber
\\
\bar{\Psi}\gamma ^\mu \Lambda &=&\psi \sigma ^\mu \bar{\lambda}+\bar{\psi}%
\bar{\sigma}^\mu \lambda \mbox{ .}  \label{1.4a}
\end{eqnarray}

We have not considered coupling with matter fields and a propagation term
for the gauge fields. On the other hand, our superspace BF term was
constructed in a very simple way. A quite similar construction was
introduced by Clark {\it et al.} \cite{Clark:1989gx}.

The off-diagonal mass term $\xi \lambda $ (or $\bar{\Xi}\gamma ^5\Lambda $)
has been shown by Brooks and Gates, Jr. \cite{Brooks:1994nn} in the context
of super-Yang-Mills theory. Note that the identity
\begin{equation}
\gamma _5\sigma ^{\mu \nu }=\frac i2\varepsilon _{\mu \nu \alpha \beta
}\sigma ^{\alpha \beta }  \label{civita}
\end{equation}
reveals a connection between the topological behaviour denoted by the
Levi-Civita tensor $\varepsilon _{\mu \nu \alpha \beta },$ and the
pseudo-scalar $\gamma _5.$

So, it is worthwhile to mention that this term has topological origin and it
can be seen as a fermionic counterpart of the BF term. In our opinion, this
fermionic mass term deserves more attention.

\vskip0.3cm

\begin{itemize}
\item  {\bf {The $N=2-D=3$ massive $B\wedge \varphi $ model}}
\end{itemize}

\vskip0.3cm

We will now carry out a dimensional reduction in the bosonic sector of (\ref
{comp}). Hence, after dimensional reduction, the bosonic sector of (\ref
{comp}) can be written as \cite{Gomes:2001fr}
\begin{eqnarray}
S_{bos.} &=&\int d^3x\{[\kappa \varepsilon _{\mu \alpha \beta }V^\mu
F^{\alpha \beta }+\kappa \varepsilon _{\mu \nu \alpha }B^{\mu \nu }\partial
^\alpha \varphi -\kappa DN]  \nonumber \\
&&+g^2[\frac 12B^{\mu \nu }B_{\mu \nu }-V^\mu V_\mu -\frac 12\left(
M^2+N^2\right) ]\}\mbox{ },  \label{1.5}
\end{eqnarray}
where $V^\mu $ is a vectorial field and $\varphi $ represents a real scalar
field. Notice that the first term in r.h.s. of (\ref{1.5}) can be
transformed in the Chern-Simons term if we identify $V^\mu \equiv A^\mu $.

The second one is the so called $B\wedge \varphi $ term.

Now let us proceed to the dimensional reduction of the fermionic sector of
the model. First, note that the Lorentz group in three dimensions is $%
SL(2,R) $ rather than $SL(2,C)$ in $D=4$. Therefore, Weyl spinors with four
degrees of freedom will be mapped into Dirac spinors. So the correct
associations keeping the degrees of freedom are sketched as
\begin{eqnarray}
\Xi &=&\left(
\begin{array}{c}
\xi _\alpha \\
\bar{\xi}^{\dot{\alpha}}
\end{array}
\right) \rightarrow \Xi _{\pm }=\xi _\alpha \pm i\tau _\alpha  \nonumber \\
\Lambda &=&\left(
\begin{array}{c}
\lambda _\alpha \\
\bar{\lambda}^{\dot{\alpha}}
\end{array}
\right) \rightarrow \Lambda _{\pm }=\lambda _\alpha \pm i\rho _\alpha
\nonumber \\
\Psi &=&\left(
\begin{array}{c}
\psi _\alpha \\
\bar{\psi}^{\dot{\alpha}}
\end{array}
\right) \rightarrow \Psi _{\pm }=\psi _\alpha \pm i\chi _\alpha \mbox{ }.
\label{1.6}
\end{eqnarray}
From ($\ref{1.6}$), we find that
\begin{eqnarray}
\Psi \bar{\Xi} &\rightarrow &\frac 12\left( \Psi _{+}\Xi _{-}+\Psi _{-}\Xi
_{+}\right)  \nonumber \\
\bar{\Psi}\gamma ^\mu \partial _\mu \Lambda &\rightarrow &\frac 12(\Psi
_{+}\gamma ^{\widehat{\mu }}\partial _{\widehat{\mu }}\Lambda _{-}+\Psi
_{-}\gamma ^{\widehat{\mu }}\partial _{\widehat{\mu }}\Lambda _{+})
\nonumber \\
\Xi \gamma ^5\Lambda &\rightarrow &\frac 12(\Xi _{+}\Lambda _{+}+\Xi
_{-}\Lambda _{-})\mbox{ }.  \label{1.7}
\end{eqnarray}
where $hatted$ index means three-dimensional space-time.

Thus, the dimensionally reduced fermionic sector of (\ref{comp}) may be
written
\begin{eqnarray}
S_{ferm.} &=&\int d^3x\{\frac{i\kappa }4\left( \Xi _{+}\Lambda _{+}+\Xi
_{-}\Lambda _{-}\right) +\frac \kappa 4(\Psi _{+}\gamma ^{\widehat{\mu }%
}\partial _{\widehat{\mu }}\Lambda _{-}  \nonumber \\
&&+\Psi _{-}\gamma ^{\widehat{\mu }}\partial _{\widehat{\mu }}\Lambda _{+})+%
\frac{g^2}{16}\left( \Psi _{+}\Xi _{-}+\Psi _{-}\Xi _{+}\right) \}\mbox{ .}
\label{1.8}
\end{eqnarray}

The action $S=S_{bos.}+S_{ferm.}$ is invariant under the following
supersymmetry transformations (from now on, Greek indices mean
three-dimensional space-time):
\begin{eqnarray}
\delta \lambda _\alpha &=&-iD\eta _\alpha -\left( \sigma ^\mu \sigma ^\nu
\right) _\alpha ^\beta \eta _\beta F_{\mu \nu }  \nonumber \\
\delta \rho _\alpha &=&iD\zeta _\alpha -\left( \sigma ^\mu \sigma ^\nu
\right) _\alpha ^\beta \zeta _\beta F_{\mu \nu }  \nonumber \\
\delta F^{\mu \nu } &=&i\partial ^\mu \left( \eta \sigma ^\nu \rho -\lambda
\sigma ^\nu \zeta \right) -i\partial ^\nu \left( \eta \sigma ^\mu \rho
-\lambda \sigma ^\mu \zeta \right)  \nonumber \\
\delta D &=&\partial _\mu \left( -\eta \sigma ^\mu \rho +\lambda \sigma ^\mu
\zeta \right)  \label{1.9a}
\end{eqnarray}
\begin{eqnarray}
\delta \left( \psi _\alpha \pm i\chi _\alpha \right) &=&\delta \Psi _{\pm
}=i\eta ^\beta \widetilde{T}_{\beta \alpha }\pm \zeta ^\beta \widetilde{T}%
_{\beta \alpha } ,  \nonumber \\
\delta \widetilde{T}_{\beta \alpha } &=&-\eta _\beta \xi _\alpha +\zeta
^\lambda \sigma _{\beta \lambda }^\mu \partial _\mu \psi _\alpha ,  \nonumber
\\
\delta \left( \xi _\alpha \pm i\tau _\alpha \right) &=&\delta \Xi _{\pm
}=-i\zeta _\lambda \left( \sigma ^\mu \right) ^{\lambda \beta }T_{\beta
\alpha }  \nonumber \\
\mp \eta _\lambda \left( \bar{\sigma}^\mu \right) ^{\beta \lambda }T_{\beta
\alpha }\mbox{
,}  \label{1.9b}
\end{eqnarray}
where $\eta $ and $\zeta $ are supersymmetric parameters, which indicates
that we have two supersymmetries in the aforementioned action.

\section{FRACTIONAL STATISTICS - ANYONS}

The fractional statistics \cite{Wilczek:1982rv} with its theoretical and
applicable consequences plays an interesting interplay role between quantum
field theory and condensed matter physics. Previous speculations \cite
{Laughlin:1983fy} that the fractional quantum Hall effect could be explained
by quasiparticles (anyons) obeying fractional statistics were confirmed and
the behaviour of two-dimensional materials such as vortices in superfluid
helium films may be explained by fractional statistics. As it is known, the
presence of Chern-Simons terms in (2+1) dimensional gauge theories induce
fractional statistics. In such theories, it has been known that there exist
excitations, called anyons, which continuously interpolate between bosons
and fermions. In the well-known physical realization, anyons are composite
quasi-particles where magnetic flux-tubes are attached to charged particles.

Recently, there have been thoughts of generalizing exotic statistics to
extended objects to the case of strings in four dimensions \cite
{Bergeron:1995ym}. Abelian BF models in four dimensions has also been
exploited in dual models of cosmic strings, and axionic black hole theories
where the axion charge is physically detectable only by external cosmic
strings in a four dimensional Aharanov-Bohm type process \cite{Alford:1989sj}%
. Aneziris {\it et al.} \cite{Aneziris:1991gm} showed that more general
statistics can exist in (3+1) dimensions. Statistical phases of BF theory
can be seen to arise from certain cosmic string and superstring phenomena,
as well as in the Nambu-Goto string theory modified with the inclusion of
the Kalb-Ramond field ( B field) \cite{Rohm:1986jv}.

\vskip1cm

\section{LINKING NUMBER - INTERSECTION NUMBER}

In a recent interesting work, Ashtekar and Corichi \cite{Ashtekar:1997rg}
showed that there is a precise in which the Heisenberg uncertainty between
fluxes of electric and magnetic fields through finite surfaces is given by
the Gauss linking number of the loops that bound these surfaces.

Topological field theories presents observables other than the partition
function. Witten has argued that in these theories Wilson loops are
appropriate metric independent and gauge invariant objects. Polyakov has
related the vacuum expectation values of Wilson loops in the abelian
Chern-Simons theory to the classical Gauss linking number of two loops.

In the case of BF systems, we can reinterpreting the linking number as the
intersection number of one loop with a disc bounded by the other loop. So,
this observable has a natural generalization to other dimensions.
Considering the action (\ref{BFD}), the fields $B_{p}$ and $A_{D-p-1}$ allow
us to form the following metric independent and gauge invariant expressions
(''Wilson surfaces''):
\begin{equation}
W[L]=\exp \left( \int_LA\right) , W[{\mit\Sigma }]=\exp \left( \int_{\mit%
\Sigma }B\right)  \label{b0}
\end{equation}
where $\Sigma $ and $L$ are disjoint compact and oriented $p-$ and $(D-p-1)$%
-dimensional boundaries of two oriented submanifold of an $D$-dimensional
oriented manifold ${\cal M}.$ This formalism was presented by Blau and
Thompson \cite{Blau:1991bq}, who proved that the expectation value $W(\Sigma
,L)=\left\langle W_B(\Sigma )W_A(L)\right\rangle $ is equal to the linking
number of the ''surfaces ''.


\vskip1cm

\section{FRACTIONAL STATISTICS IN D=3 FROM $B\wedge \varphi $ TERM?}

Consider the following action
\begin{equation}
S=S_0+\int d^3x\left( \frac \kappa 2\varepsilon ^{\mu \nu \alpha }B_{\mu \nu
}\partial _\alpha \varphi +\frac g2J^{\mu \nu }B_{\mu \nu }+hj\varphi
\right) ,  \label{b1}
\end{equation}
where $g,h$ are coupling constants, $J^{\mu \nu }$ and $j$ are currents and
sources. $S_0$ depends only on fields that originate currents and sources.
Integrating out the fields $B_{\mu \nu }$ and $\varphi $, we arrive at
\begin{equation}
S_{eff}=S_0-\frac{hg}{4\kappa }\int \int d^3xd^3yJ^{\mu \nu }(x)\left\langle
B_{\mu \nu }(x)\varphi (y)\right\rangle j(y).  \label{b2}
\end{equation}

From (\ref{b1}) and using the Landau gauge, is easy to see that
\begin{equation}
\langle B_{\mu \nu }(x)\varphi (x)\rangle =\varepsilon _{\mu \nu \alpha
}\partial _x^\alpha G(x-y),  \label{p1}
\end{equation}
where
\begin{equation}
G(x-y)=-\frac 1{4\pi }\frac 1{\left| x-y\right| },  \label{p2}
\end{equation}
Therefore
\begin{equation}
\langle B_{\mu \nu }(x)\varphi (x)\rangle =\frac{\varepsilon _{\mu \nu
\alpha }}{4\pi }\frac{(x-y)^\alpha }{\left| x-y\right| ^3},  \label{p3}
\end{equation}
The correlation function $\left\langle B_{\mu \nu }(x)\varphi
(y)\right\rangle $ is tantamount to the correlation function $\left\langle
A_\mu (x)A_\nu (y)\right\rangle $ of the pure Chern-Simons theory in the
Landau gauge (transverse propagator). The effective action (\ref{b2}) can be
rewritten as
\begin{equation}
S_{eff}=S_0-\frac{hg}{4\kappa }\frac 1{4\pi }\varepsilon _{\mu \nu \alpha
}\int \int d^3xd^3yJ^{\mu \nu }(x)\frac{(x-y)^\alpha }{\left| x-y\right| ^3}%
j(y)  \label{b3}
\end{equation}
and
\begin{equation}
S_{eff}=S_0-\frac{hg}{4\kappa }({linking \hskip0.3cm number})  \label{b3.1}
\end{equation}
On the other hand, Blau and Thompson \cite{Blau:1991bq} suggest application
of their formalism to the case where $B$ is a zero-form and $A$ is a
one-form, involving a linking number of a point $P$ and a circle $\gamma $,
through the expression
\begin{equation}
W_B(P)W_A(d)=\exp (B(P)+\oint_\gamma A)  \label{b4}
\end{equation}
where a disc $d$ is bounded by $\gamma $. These results support our
speculation that we can define a linking number from the $B\wedge \varphi $
term, and that it can exist a fractional statistics even in this case.


\vskip1cm

\section{PAULI 'S TERM AND FRACTIONAL STATISTICS IN D=3.}

As it is known, the presence of Chern-Simons terms in (2+1) dimensional
gauge theories induce fractional statistics\cite
{Semenoff:1988jr,Dunne:1994uy}. Stern \cite{Stern:1991fm} was the first, as
far as we know, to suggest a nonminimal term in the context of the
Maxwell-Chern-Simons electrodynamics with the intention of mimicking an
anyonic behaviour without a pure Chern-Simons limit. This term can be
interpreted as a tree level Pauli-type coupling, {\it i. e.}, an anomalous
magnetic moment. It is a specific feature of (2+1) dimensions that the Pauli
coupling exists, not only for spinning particles, but also for scalar ones
\cite{Torres:1992wz}. We consider here an Abelian Chern-Simons-Higgs theory
where the complex scalar fields couples directly to the electromagnetic
field strength (Pauli-type coupling). The Lagrangian of the model under
investigation is
\begin{equation}
L=\left| \nabla _\mu \phi \right| ^2+\frac \kappa 2\varepsilon ^{\mu \nu
\lambda }A_\mu \partial _\nu A_\lambda -A_\mu \partial ^\mu b+\frac \alpha 2%
b^2  \label{eq.1}
\end{equation}
where $\nabla _\mu \phi \equiv (\partial _\mu -ieA_\mu -i\frac g4\varepsilon
_{\mu \lambda \sigma }F^{\lambda \sigma })\phi $. Note that this covariant
derivative includes both the usual minimal coupling and the contribution due
to Pauli's term. Here $A_\mu $ is the gauge field and the Levi-Civita symbol
$\varepsilon _{\mu \nu \lambda }$ is fixed by $\varepsilon _{012}=1$ and $%
g_{\mu \nu }=diag(1,-1,-1)$. The multiplier field $b$ has been introduced to
implement the covariant gauge-fixing condition.

Before quantizing the theory, we analyze the above Lagrangian in terms of
Hamiltonian methods. Here we follow the approach used by Shin {\it et al.}
\cite{Shin:1992ar}. We carry out the constraint analysis of this model, in
order to obtain a consistent formulation of the theory.

The canonical momenta of the Lagrangian (\ref{eq.1}), which can be easily
seen by considering its temporal and spatial components separately, are
given by
\begin{equation}
\pi _0=0,  \label{eq.2a}
\end{equation}
\begin{equation}
\pi _b=-A_0,  \label{eq.2b}
\end{equation}
\begin{eqnarray}
\pi ^j=-\frac \kappa 2\varepsilon ^{ij}A_i-i\frac g2\varepsilon ^{ij}\left[
\phi ^{*}(D_i\phi )-\phi (D_i\phi )^{*}\right]  \nonumber \\
-\frac{g^2}2\partial ^jA_0\left| \phi \right| ^2+\frac{g^2}4(\partial
_0A_0)\left| \phi \right| ^2 ,  \label{eq.2c}
\end{eqnarray}
\begin{equation}
\pi =(\partial _0\phi ^{*})+ieA_0\phi ^{*}+i\frac g4\phi ^{*}\varepsilon
^{ij}F_{ij},  \label{eq.2d}
\end{equation}
\begin{equation}
\pi ^{*}=(\partial _0\phi )-ieA_0\phi -i\frac g4\phi \varepsilon ^{ij}F_{ij}
,  \label{eq.2e}
\end{equation}
where $\pi _0$, $\pi ^j,$ $\pi _{b},$ $\pi $ and $\pi ^{*}$ are the
canonical momenta conjugate to $A_0,$ $A_j$, $b,$ $\phi $ and $\phi ^{*}$
respectively. Also we have used $\varepsilon _{ij}=\varepsilon _{0ij}$ , $%
D_i=$ $\partial _i-ieA_{i}$ and $i,j=1,2$ .

The canonical momenta (\ref{eq.2a}) and (\ref{eq.2b}) do not involve
explicit time dependence and hence are primary constraints. Performing the
Legendre transformation, the canonical Hamiltonian can be written as
\begin{eqnarray}
H_c=\pi ^{*}\pi +\left| D\phi \right| ^2+ieA_0(\pi \phi -\pi ^{*}\phi
^{*})+\kappa \varepsilon ^{ij}A_0\partial _iA_j  \nonumber \\
+A_i\partial ^ib-\frac \alpha 2b^2-i\frac g2\varepsilon ^{ij}\partial
_jA_0\left[ \phi ^{*}(D_i\phi )-\phi (D_i\phi )^{*}\right]  \nonumber \\
-\frac{g^2}4\partial _iA_0\partial ^iA_0\left| \phi \right| ^2-\frac g4%
\varepsilon ^{ij}F_{ij}\left[ \phi ^{*}(D_0\phi )-\phi (D_0\phi )^{*}\right]
\nonumber \\
-\frac{g^2}8F^{ij}F_{ij}\left| \phi \right| ^2.  \label{eq.3}
\end{eqnarray}

Now, in order to implement the primary constraints in the theory, we
construct the primary Hamiltonian as
\begin{equation}
H_p=H_c+\lambda _0\pi +\lambda _1(\pi _b+A_0),  \label{eq.4}
\end{equation}
where $\lambda _0$ and $\lambda _1$ are Lagrange multiplier fields.
Conserving in time the primary constraints yields the secondary constraints
\begin{equation}
\psi _1=\pi _0\approx 0,  \label{eq.5a}
\end{equation}
\begin{equation}
\psi _2=\pi _b+A_0\approx 0,  \label{eq.5b}
\end{equation}
which are also conserved in time and where the symbol $\approx $ indicates
weak equality, {\it i. e.}, the constraints can be identically set equal to
zero only after computing the relevant Poisson brackets. Thus there is no
more constraint and the above equations are the set of fully second-class
constraints. On the other hand, there is no first-class conditions and so,
no gauge conditions to be determined in theory. This is an effect of the
gauge fixing condition imposed previously. As it is known, the lack of
physical significance allows that the second-class constraints can be
eliminated by means of Dirac brackets (DB's).

Following the standard Dirac brackets formalism and quantizing the system,
we obtain the following set of non-vanishing equal-time commutators:
\begin{equation}
\left[ A_0(x),b(y)\right] =i\delta ^2(x-y)  \label{eq.6a}
\end{equation}
\begin{equation}
\left[ A_i(x),\pi _j(y)\right] =i\delta _{ij}\delta ^2(x-y)  \label{eq.6b}
\end{equation}
\begin{equation}
\left[ \phi (x),\pi (y)\right] =\left[ \phi ^{*}(x),\pi ^{*}(y)\right]
=i\delta ^2(x-y)  \label{eq.6c}
\end{equation}
After achieving the quantization we proceed to construct the angular
momentum operator and compute the angular momentum of the matter field $\phi
$.

The symmetric energy-momentum tensor can be obtained by coupling the fields
to gravity and then varying the action with respect to $g^{\mu \nu }$:
\begin{eqnarray}
T_{\mu \nu }=\frac 2{\sqrt{-g}}\frac{\delta S}{\delta g^{\mu \nu }} =(\nabla
_\mu \phi )^{*}(\nabla _\nu \phi )+(\nabla _\nu \phi )^{*}(\nabla _\mu \phi )
\nonumber \\
-A_\mu \partial _\nu b-A_\nu \partial _\mu b  \nonumber \\
-g_{\mu \nu }(\left| \nabla _\alpha \phi \right| ^2-A_\alpha \partial
^\alpha b).  \label{eq.7}
\end{eqnarray}
The angular momentum operator in (2+1) dimensions is given by
\[
L=\int d^2x\varepsilon ^{ij}x_iT_{0j}.
\]
Hence
\begin{eqnarray}
L =\int d^2x\varepsilon ^{ij}x_i\{(\pi \partial _j\phi +\pi ^{*}\partial
_j\phi ^{*})-ieA_jJ_0  \nonumber \\
-i\frac g2\varepsilon _{jl}F^{l0}(\pi \phi -\pi ^{*}\phi ^{*})-A_0\partial
_jb  \nonumber \\
+A_j\partial _0b-i\frac g2A_j\varepsilon ^{kl}\partial _k[\phi ^{*}(D_l\phi
)-\phi (D_l\phi )^{*}]  \nonumber \\
+i\frac{g^2}2A_j\partial _k(\left| \phi \right| ^2F^{0k})\},  \label{eq.8}
\end{eqnarray}
where
\begin{eqnarray}
J_0=i\{\pi \phi -\pi ^{*}\phi ^{*}-\frac g{2e}\varepsilon ^{ij}\partial
_i[\phi ^{*}(D_j\phi )-\phi (D_j\phi )^{*}]  \nonumber \\
+i\frac{g^2}{2e}\partial _i(\left| \phi \right| ^2F^{0i})\}  \label{eq.9}
\end{eqnarray}
is the temporal component of the conserved matter current. The key point
here is that Gauss' law is no more a constraint, while $J_0$ and $T_{\mu \nu
}$ contain derivatives of $A_\mu $ . Note that, due to its topological
character, the Chern-Simons term does not contribute to the energy-momentum
tensor. These aspects are attributed to the non-linearity introduced by
Pauli's term.

The rotational property of the $\phi $ field is obtained by computing the
commutator $[L,\phi (y)]$. Using equations (\ref{eq.6a}-\ref{eq.6c}) and (%
\ref{eq.8}), it is easy to see that
\begin{eqnarray}
\lbrack L,\phi (y)]=\varepsilon ^{ij}y_i\partial _j\phi -[e\int
d^2x\varepsilon ^{ij}x_iA_jJ_0,\phi ]  \nonumber \\
+i\frac g2\varepsilon ^{ij}\varepsilon_{jk}y_iF^{k0}\phi.  \label{eq.10}
\end{eqnarray}
This commutator can be rewritten by means of the electromagnetic charge
operator
\[
Q=\int d^2xJ_0(x)
\]
and becomes
\begin{equation}
\lbrack L,\phi (y)]=\varepsilon ^{ij}y_i\partial _j\phi -\frac{e^2}{4\pi
\kappa }[Q^2,\phi (y)]+i\frac g2\varepsilon ^{ij}\varepsilon
_{jk}y_iF^{k0}\phi  \label{eq.11}
\end{equation}
or, in more familiar notation
\begin{equation}
\lbrack L,\phi (y)]=i({\bf y}\times {\bf \nabla })\phi (y)-\frac{e^2}{2\pi
\kappa }Q\phi (y)+i\frac g2{\bf y\cdot E}\phi (y).  \label{eq.12}
\end{equation}

The first term in the right hand side of eq. (\ref{eq.12}) represents the
intrinsic spin and the second is the so-called rotational anomaly, which is
responsible for the fractional spin. Unlike the Chern-Simons term (whose
contribution is related with magnetic field), the Pauli term induces an
anomalous contribution for the spin of the system, which depends on electric
field \cite{Nobre:1999mj}. We stress that, here the nonminimal coupling
constant is a free parameter.

It is worth mentioning that all the procedure above can be carried out even
if there is no Chern-Simons term in the Lagrangian (\ref{eq.1}). In this
case the anomalous contribution to spin would just come from the Pauli term.

Now we will discuss the above result in connection with theories in the
broken-symmetry phase. Boyanovsky \cite{Boyanovsky:1990ks} has found that
the low-lying excitations of a U(1) Chern-Simons theory in interaction with
a complex scalar field in a broken symmetry state are massive bosons with
{\it canonical statistics}. He explained his result as due to the screening
of long-range forces in a broken symmetry phase. In this phase localized
charge distributions cannot be supported, which is supposed to be essential
for fractional spin. On the other hand, if we consider a non-minimally
coupled Abelian-Higgs model, the long-distance damping effect by the
''photon'' mass $\kappa $ no longer exists. This is an indication that
Pauli's term, which induces an anomalous\- spin, can be relevant for the
study of broken symmetry states (superfluid) in the context of effective
theories in condensed matter.

In nonrelativistic limit, Carrington and Kunstatter \cite{Carrington:1995km}
have shown that anomalous magnetic moment interactions gives rise to both
the Aharonov-Bohm and Aharonov-Casher effects. They have speculated possible
anomalous statistics without the CS term. As a matter of fact, we believe
that this (in a relativistic theory) was proved here. On the other hand, the
Abelian Chern-Simons term can be generated by means of a spontaneous
symmetry breaking of a nonminimal theory. This connection between
Chern-Simons and Pauli-type coupling was pointed out by Stern. So the Pauli
term at tree-level (with the nonminimal coupling constant $g$ as a free
parameter) can constitute an effective theory which bring us information
about physical models in broken symmetry phase.

\section{ACKNOWLEDGMENTS}

I would like to thank my collaborators R. R. Landim, D. M. Medeiros, F. A.
S. Nobre and M. A. M. Gomes which contributed for many results reported
here. This work was supported in part by Conselho Nacional de
Desenvolvimento Cient\'{\i }fico e Tecnol\'{o}gico-CNPq and Funda\c{c}\~{a}o
Cearense de Amparo \`{a} Pesquisa-FUNCAP.


\begin{references}


\bibitem{Birmingham:1991ty}  D.~Birmingham, M.~Blau, M.~Rakowski and
G.~Thompson, 
Phys.\ Rept.\ {\bf 209}, 129 (1991). 


\bibitem{Allen:1991gb}  T.~J.~Allen, M.~J.~Bowick and A.~Lahiri,
Mod.\ Phys.\ Lett.\ A {\bf 6}, 559 (1991). 


\bibitem{Medeiros:1999df}  D.~M.~Medeiros, R.~R.~Landim and C.~A.~Almeida,
Europhys.\ Lett.\ {\bf 48}, 610 (1999) [hep-th/9906124].


\bibitem{Fucito:1997ax}  F.~Fucito, M.~Martellini and M.~Zeni,
Nucl.\ Phys.\ B {\bf 496}, 259 (1997) [hep-th/9605018].


\bibitem{Gomes:2001fr}  M.~A.~Gomes, R.~R.~Landim and C.~A.~Almeida,
Phys.\ Rev.\ D {\bf 63}, 025005 (2001) [hep-th/0005004].


\bibitem{Clark:1989gx}  T.~E.~Clark, C.~H.~Lee and S.~T.~Love,
Mod.\ Phys.\ Lett.\ A {\bf 4}, 1343 (1989). 


\bibitem{Brooks:1994nn}  R.~Brooks and S.~J.~Gates,
Nucl.\ Phys.\ B {\bf 432}, 205 (1994) [hep-th/9407147].


\bibitem{Wilczek:1982rv}  F.~Wilczek,
Phys.\ Rev.\ Lett.\ {\bf 49}, 1549 (1982). 


\bibitem{Laughlin:1983fy}  R.~B.~Laughlin,
Phys.\ Rev.\ Lett.\ {\bf 50}, 1395 (1983). 


\bibitem{Bergeron:1995ym}  M.~Bergeron, G.~W.~Semenoff and R.~J.~Szabo,
Nucl.\ Phys.\ B {\bf 437}, 695 (1995) [hep-th/9407020].


\bibitem{Alford:1989sj}  M.~G.~Alford and F.~Wilczek,
Phys.\ Rev.\ Lett.\ {\bf 62}, 1071 (1989). 


\bibitem{Aneziris:1991gm}  C.~Aneziris, A.~P.~Balachandran, L.~Kauffman and
A.~M.~Srivastava,
Int.\ J.\ Mod.\ Phys.\ A {\bf 6}, 2519 (1991).


\bibitem{Rohm:1986jv}  R.~Rohm and E.~Witten,
Annals Phys.\ {\bf 170}, 454 (1986). 


\bibitem{Ashtekar:1997rg}  A.~Ashtekar and A.~Corichi,
Phys.\ Rev.\ D {\bf 56}, 2073 (1997) [hep-th/9701136].


\bibitem{Blau:1991bq}  M.~Blau and G.~Thompson,
Annals Phys.\ {\bf 205}, 130 (1991). 


\bibitem{Semenoff:1988jr}  G.~W.~Semenoff,
Phys.\ Rev.\ Lett.\ {\bf 61}, 517 (1988). 


\bibitem{Dunne:1994uy}  G.~Dunne, ``Selfdual Chern-Simons theories,''
hep-th/9410065. 


\bibitem{Stern:1991fm}  J.~Stern,
Phys.\ Lett.\ B {\bf 265}, 119 (1991). 


\bibitem{Torres:1992wz}  M.~Torres,
Phys.\ Rev.\ D {\bf 46}, 2295 (1992). 


\bibitem{Shin:1992ar}  H.~Shin, W.~Kim, J.~Kim and Y.~Park,
Phys.\ Rev.\ D {\bf 46}, 2730 (1992). 


\bibitem{Nobre:1999mj}  F.~A.~Nobre and C.~A.~Almeida,
Phys.\ Lett.\ B {\bf 455}, 213 (1999) [hep-th/9904159].


\bibitem{Boyanovsky:1990ks}  D.~Boyanovsky,
Phys.\ Rev.\ D {\bf 42}, 1179 (1990). 


\bibitem{Carrington:1995km}  M.~E.~Carrington and G.~Kunstatter,
Phys.\ Rev.\ D {\bf 51}, 1903 (1995). 
\end{references}
\end{document}